\documentclass[twocolumn,tighten]{aastex62}
\hypersetup{linkcolor=red,citecolor=blue,filecolor=cyan,urlcolor=magenta}
\usepackage{amsmath}
\usepackage{natbib}
\usepackage{color}
\usepackage{listings}

\newcommand{\msun}{\mbox{$M_{\odot}$}}

\newcommand{\pc}{\ensuremath{\,\mathrm{pc}}}
\newcommand{\kpc}{\ensuremath{\,\mathrm{kpc}}}
\newcommand{\Myr}{\ensuremath{\,\mathrm{Myr}}}
\newcommand{\Gyr}{\ensuremath{\,\mathrm{Gyr}}}
\newcommand{\kms}{\ensuremath{\,\mathrm{km\ s}^{-1}}}

\newcommand{\mas}{\ensuremath{\,\mathrm{mas}}}
\newcommand{\masyr}{\ensuremath{\,\mathrm{mas\ yr}^{-1}}}

\newcommand{\vlos}{v_{\ensuremath{\mathrm{los}}}}
\newcommand{\muell}{\mu_{\ell*}}
\newcommand{\mub}{\mu_{b}}
\newcommand{\mualpha}{\mu_{\alpha*}}
\newcommand{\mudelta}{\mu_\delta}

\newcommand{\eq}[1]{\begin{align}#1\end{align}}

\shorttitle{Extreme velocity stars in Gaia DR2}
\shortauthors{Hattori et al.}
\accepted{for publication in ApJ}

\begin{document}

\title{%
Old, Metal-Poor Extreme Velocity Stars in the Solar Neighborhood\footnote{Based on data from the \textit{Gaia}-DR2 Archive }
}

\author{Kohei Hattori}
\affiliation{%
Department of Astronomy, University of Michigan,
1085 S.\ University Ave., Ann Arbor, MI 48109, USA}
\email{Email:\ khattori@umich.edu}

\author{Monica Valluri}
\affiliation{%
Department of Astronomy, University of Michigan,
1085 S.\ University Ave., Ann Arbor, MI 48109, USA}

\author{Eric F. Bell}
\affiliation{%
Department of Astronomy, University of Michigan,
1085 S.\ University Ave., Ann Arbor, MI 48109, USA}

\author{Ian U.\ Roederer}
\affiliation{%
Department of Astronomy, University of Michigan,
1085 S.\ University Ave., Ann Arbor, MI 48109, USA}
\affiliation{%
Joint Institute for Nuclear Astrophysics -- Center for the
Evolution of the Elements (JINA-CEE), USA}

\begin{abstract}
We report the discovery of 30 stars  
with extreme space velocities ($\gtrsim 480 \kms$) in the {\it Gaia}-DR2 archive. 
These 
stars are a subset of 1743 stars with high-precision parallax, large tangential velocity ($v_{\rm tan}>300 \kms$), and measured line-of-sight velocity in DR2. 
By tracing the orbits of the stars back in time, we find at least one of them is consistent with having been ejected by the supermassive black hole at the Galactic Center. 
Another star has an orbit that passed near the Large Magellanic Cloud (LMC) about 200 Myr ago. 
Unlike previously discovered blue hypervelocity stars, our sample is metal-poor ($-1.5 <$~[Fe/H]~$< -1.0$) and quite old ($>1 \Gyr$). 
We discuss possible mechanisms for accelerating old stars to such extreme velocities. 
The high observed space density of this population, relative to potential acceleration mechanisms, implies that these stars are probably bound to the Milky Way (MW). If they are bound, the discovery of this population would require a local escape speed of around $\sim 600 \kms$ and consequently imply a virial mass of $M_{200} \sim 1.4 \times 10^{12} M_\odot$ for the MW. 
\end{abstract}

\section{Introduction}
\label{intro}

The origin of unbound stars with extremely large velocities in the halo of our Milky Way (MW) is currently unknown. 
So far, about 20 ``hypervelocity stars'' (HVSs) with velocities above $400$-$500 \kms$ have been identified in the distant halo, and most stars are confirmed to be young, massive stars, such as B-type main sequence stars \citep[e.g.,][and references therein]{Brown2015ARAA}. 
Because they are typically located over $50 \kpc$ from star formation sites, these young stars are believed to be recently ejected from some star forming regions near the Galactic Center, the MW stellar disk, or star-forming dwarf satellites of the Galaxy.

The most widely recognized mechanism to eject a star with a large velocity is associated with the supermassive black hole (SMBH) at the Galactic Center. \cite{Hills88} and \cite{YuTremaine02} theoretically proposed that the SMBH can disrupt a close binary system and eject a star with a velocity of $\sim 1000 \kms$, which allows a young star to travel to the outer halo ($\sim 50\kpc$) during its lifetime. 

Other possible ejection  mechanisms include ejection of the binary companion of a star that explodes as a supernova (SN) \citep{Blaauw61}, or the dynamical few-body interactions in young dense star clusters \citep{Leonard1991}. Both mechanisms can produce ejection velocities from the stellar disk of $\sim 600 \kms$ for main-sequence stars.

Another possibility is the ejection from star-forming dwarf galaxies. For example, the Large Magellanic Cloud (LMC), which is located at a Galactocentric radius of $r\sim 50 \kpc$, is moving at $400 \kms$ with respect to the MW; so even a relatively small ejection velocity $\sim 200 \kms$ from the LMC could produce stars with extremely large velocities in the rest frame of the MW \citep{Boubert2016}.

Different mechanisms for producing HVSs predict different observational signatures in clustering, space motions, and stellar populations. 
Precise positions and space velocities would permit backward orbit integration to potential ejection locations \citep{bromley_kenyon_09,Brown2015ApJ,Erkal2018, Marchetti2018, Brown2018, Li2018}. Access to a sample of high-velocity stars that is not restricted to young stars would be  valuable \citep{Kollmeier2010}, and would critically test ejection scenarios (e.g., Galactic Center ejection should eject metal-rich stars with a range in ages). Because the term HVSs  has primarily been used to refer to unbound stars ejected by the interaction of a stellar binary with a central black hole, we will refer to stars with large velocity as ``extreme velocity stars,'' in order to be agnostic about the acceleration mechanism. 

Local populations of stars with very high velocities are of interest for another reason: they provide candidates for measuring the local escape speed. Because the escape speed at a given radius
in the MW depends on the mass beyond that radius, it is one of the few {\em local measurements that provides constraints on the total mass of the MW}.  The current uncertainty in the mass of the MW is more than a factor of two, with values of $M_{200}$ ranging from $0.87\times 10^{12}$\msun\ \citep{Xue2008} to $2.6\times 10^{12}$\msun\ \citep{Watkins2010}. More generally, it has been found that measurements that rely on the kinematics of halo stars tend to yield systematically lower values than studies that use more distant satellites as kinematic tracers \citep{Blandhawthorn_Gerhard2016}. 
A difficulty with distant tracers is that their proper motions are unknown or highly uncertain. 
In contrast, measuring the total mass of the MW by estimating the local escape velocity ($v_{\rm esc}$) from the 3D space velocity of a local sample of stars with extremely high velocity provides a  powerful alternative. 
Previous estimates of the mass of the MW from the determination of $v_{\rm esc}$ have used line of sight velocities of stars from the RAVE survey \citep{Smith2007,Piffl2014}. The availability of a new local sample of sample of extremely high-velocity stars is therefore significant.

The {\it Gaia}-DR2 archive, in providing accurate proper motions for $>$~100 million stars and radial velocities for a subset of over 7 million of them, permits a kinematic selection of 30 stars with high space velocity (Section~\ref{sec:sample}). 
We use dust-corrected color-magnitude diagrams to characterize the sample (Section~\ref{sec:CMD}). Using \textit{Gaia} phase space coordinates for each star and assuming a popular current model potential for the MW, we attempt to determine the ejection locations of these stars by integrating their orbits back in time (Section \ref{sec:orbits}). 
We discuss possible acceleration mechanisms and the implications for the local escape velocity and the mass of the MW in Section~\ref{sec:discussion}.

\section{Sample Selection and Orbit Computation}
\label{sec:sample}

{\it Gaia} DR2 includes 7,224,631 stars with line-of-sight velocities ($\vlos$) obtained with the {\it Gaia} Radial Velocity Spectrometer \citep{2018arXiv180409365G, Katz2018}. 
In this paper, we first select stars based on their high tangential velocities, and then use  $\vlos$ from {\it Gaia} to compute  the total velocity $v_{\rm total}$ for each star, and thereby select the final sample.  
Therefore, by design, the total velocity $v_{\rm total}$ of our sample stars is not dominated by $\vlos$. Thus, any  errors in the measurement of $\vlos$ should not seriously affect the number of extreme velocity stars. For a subset of stars with $\vlos$ measurements from LAMOST or RAVE, the reported values are quite consistent with Gaia RVS measurements.

We first identified 1743 candidate stars from the {\it Gaia}-DR2 archive that have:\
(i) high-precision parallax ($\varpi / \delta \varpi>10$) implying a distance accuracy of better than 10\% (ensuring accurate tangential velocity measurements);
(ii) measured $\vlos$; and 
(iii) $v_{\rm tan} > 300 \kms$.
Here, $v_{\rm tan}$ is the Galactic rest frame tangential velocity 
corrected for the solar reflex motion, and is given by: 
\eq{
&v_{\rm tan}^2 = 
[k \muell / \varpi - U_\odot \sin \ell + V_\odot \cos \ell ]^2 +\nonumber \\
&
[ k \mub / \varpi - U_\odot \cos \ell \sin b - V_\odot \sin \ell \sin b + W_\odot \cos b ]^2 , 
}
where $k=4.74047 \kms\kpc^{-1} {(\masyr)}^{-1}$,
$\varpi$ is the parallax, 
$(\ell, b)$ are the Galactic longitude and latitude,
$(\muell, \mub)$ are the associated proper motion components, and 
$(U_\odot, V_\odot, W_\odot) = (11.1, 232.24, 7.25) \kms$ 
are the Galactocentric solar velocity components.
The solar peculiar velocity is taken from \citet{Schonrich2010}, 
and we assume the local standard of rest velocity of 
$v_0 = 220 \kms$. 
For these 1743 stars, 
we derive the 3D position and velocity 
in the Galactocentric rest frame 
by additionally taking into account $\vlos$ from {\it Gaia}. 
Here, we assume that the Galactocentric distance of the Sun is $R_0 = 8 \kpc$.

We assume a gravitational potential model for the MW, {\tt MWPotential2014} \citep{Bovy2015}, and evaluate the orbital energy $E$ for each star.  We use the as-observed 6D coordinates of each star and select 30 stars that are unbound ($E>0$) or marginally bound ($E>-0.1 v_0^2 = -4840 \;\mathrm{km^2 \; s^{-2}}$) in this potential. All of these stars lie within $8\kpc$ of the solar position. 
For each of these 30 stars, we use Monte Carlo sampling to draw 1000 current positions and velocities from the error distribution around the observed 6D 
quantities, by fully taking into account the correlations in the error. 
Next, we evaluate the probability of each star being unbound, $P_\mathrm{unb}$, 
in this potential. (In Section~\ref{discussion_esc}, we discuss further the validity of this assumption.) Hereafter, we refer to these 30 stars as our ``extreme velocity'' sample, based on the fact that they are determined to be unbound in this potential, but note that at least 20 additional stars in the full sample have comparable velocities. It is important to point out that our main conclusions are unaffected by how our extreme velocity sample is selected.

\begin{figure}
\begin{center}
\includegraphics[angle=0,width=3.35in]{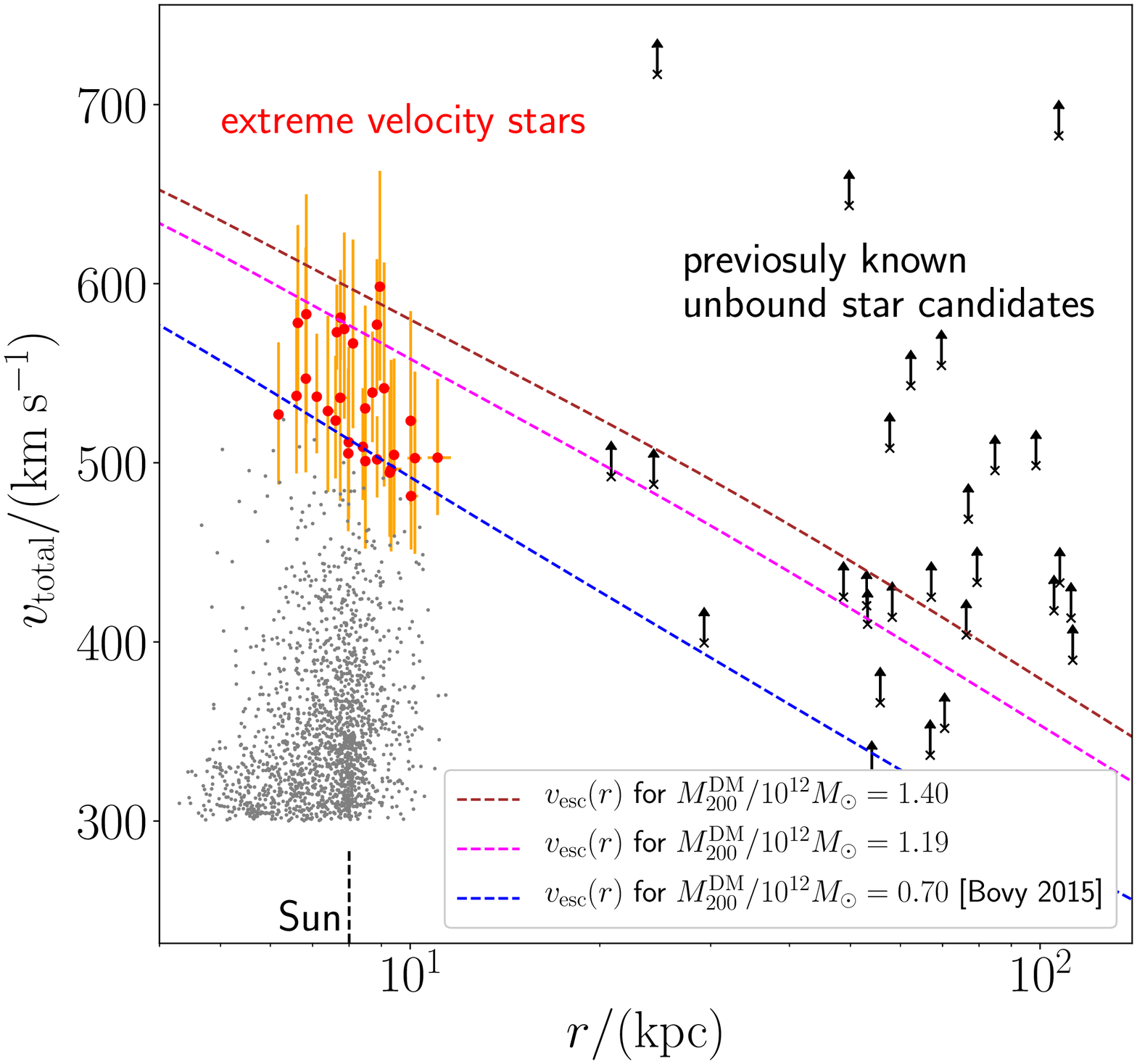} 
\end{center}
\caption{
\label{fig:r_Vtot}
The total velocity $v_{\rm total}$ and the Galactocentric radius $r$ 
for the 30 newly discovered extreme velocity stars (red dots with error bars). 
The gray dots correspond to 1713 bound stars with $v_{\rm tan}> 300 \kms$. 
Previously known hypervelocity star candidates \citep{Zheng2014, Brown2015ARAA,  Huang2017} are also shown at $r \gtrsim 20\kpc$,
with arrows marking their Galactic rest frame line-of-sight velocity used to indicate a lower bound on $v_{\rm total}$. 
The blue dashed line shows the Galactic escape velocity in the {\tt MWPotential2014} model \citep{Bovy2015}. Also plotted are the escape velocity curves for two models with higher dark halo mass.
}
\end{figure}

The sample is listed in Table~\ref{table1}, where stars are listed in descending order of their total energy $E$ in this potential, with the $i$th star in the table named Gaia-T-ES$i$ ($i=1,\cdots,30$). 
The total velocity $v_\mathrm{total}$ for these stars as a function of Galactocentric radius $r$ is shown in Figure \ref{fig:r_Vtot}.
In order to justify that the astrometric data are reasonably clean, 
we have confirmed that 
all of our final sample of 30 stars 
satisfy the flux excess criteria and the criteria on the value of 
astrometric\_chi2\_al / (astrometric\_n\_good\_obs\_al $-5$) 
as mentioned in Appendix C of \cite{Lindegren2018}.

We note that a recent paper by \cite{Marchetti2018} adopted a slightly different strategy from ours to select stars with large velocity. 
They adopted more conservative criteria for the quality of the  {\it Gaia} astrometric solution (see conditions (i)-(v) in their section 4), while they allowed large formal errors on parallax or proper motion as long as the total velocity $v_{\rm total}$ could be computed with $<$ 30\% error.  
In contrast, we select those stars with large tangential velocity and small formal error on parallax (which results in small formal error on proper motion as well),  but we do not adopt any cut on the quality of the astrometric solution. We have confirmed that the fractional error on $v_{\rm total}$ in our sample is between 4\% and 12\%. 
The differences in the strategies  adopted imply that our sample might include stars with large systematic errors on the astrometric solution (in spite of the small formal error on parallax). 
However, it is also true that their conservative cut on the quality of astrometric solution might potentially discard a lot of interesting candidate stars with small formal errors on parallax and proper motion. 
Also, they adopted a potential model for the MW, which is different from the one used here, to select stars with high probability of being unbound. 
Thus, our study is complementary to their work. 
Indeed, six stars in our final catalog (Gaia-T-ES5,6,7,10,11,15) have high-quality astrometric data and are reported in \cite{Marchetti2018}. 
Eight stars (Gaia-T-ES17,18,22,23,24,26,27,29) have high-quality astrometric data but are not reported in \cite{Marchetti2018}; 
and indeed, Gaia-T-ES22 and 29 turn out to have physically interesting orbits (see Section \ref{sec:orbits}). 
The other 16 stars have lower-quality astrometric solutions, according to their criteria. 
We expect that the quality of the data will be improved in future data releases from {\it Gaia}, so we believe that even the stars with lower-quality astrometric data in DR2 are worth  analyzing. We note that exclusion of stars with lower-quality data will not alter the main conclusions of this paper.

\begin{deluxetable*}{l r rr rr rrr rr rc}
\tablecaption{Extreme velocity star sample, sorted by decreasing orbital energy $E$
\label{table1}}
\tablewidth{0pt}
\tabletypesize{\scriptsize}
\tablehead{
\colhead{Short name} &
\colhead{GaiaDR2 source\_id} &
\colhead{$\ell$} &
\colhead{$b$} &
\colhead{$d_{\rm helio}$} &
\colhead{$r$} &
\colhead{$v_{\rm total}$} &
\colhead{$v_r$} &
\colhead{$\vlos$} &
\colhead{$E$} &
\colhead{$-L_z$} &
\colhead{$P_{\rm unb}$} &
\colhead{Origin} 
}
\startdata
           &                     & deg & deg & $\kpc$ & $\kpc$ & $\kms$ & $\kms$ & $\kms$ & $\mathrm{km^2\;s^{-2}}$ & $\kpc\kms$ &  &  \\ 
 \hline 
Gaia-T-ES1 & 3252546886080448384 & 193.87 & -36.61 & $1.18 \pm 0.09$ & 9.0 & 598.3 & -337.0 & 1.7 & 53273 & -3578 & 0.95 & $-$ \\ 
Gaia-T-ES2 & 5300505902646873088 & 278.09 & -6.83 & $5.08 \pm 0.36$ & 8.9 & 577.1 & -539.3 & 160.2 & 40228 & -1651 & 0.98 & $-$ \\ 
Gaia-T-ES3 & 2629296824480015744 & 61.28 & -46.88 & $0.89 \pm 0.04$ & 7.8 & 581.1 & 552.6 & -219.7 & 36368 & -827 & 1.00 & $-$ \\ 
Gaia-T-ES4 & 6505889848642319872 & 332.40 & -53.84 & $3.63 \pm 0.36$ & 6.8 & 583.0 & 568.9 & -38.2 & 35132 & -536 & 0.85 & $-$ \\ 
Gaia-T-ES5 & 5212817273334550016 & 287.70 & -25.27 & $3.81 \pm 0.30$ & 7.9 & 574.7 & 456.3 & 159.9 & 34520 & -28 & 0.86 & $-$ \\ 
Gaia-T-ES6 & 3705761936916676864 & 302.68 & 67.81 & $3.76 \pm 0.34$ & 8.1 & 566.7 & -524.7 & 88.7 & 33232 & -1354 & 0.86 & $-$ \\ 
Gaia-T-ES7 & 6397497209236655872 & 321.80 & -42.67 & $5.79 \pm 0.56$ & 6.6 & 578.1 & -572.7 & -8.2 & 32442 & -256 & 0.88 & $-$ \\ 
Gaia-T-ES8 & 5212110596595560192 & 289.93 & -28.26 & $2.92 \pm 0.15$ & 7.6 & 572.9 & -553.7 & 298.2 & 31964 & -842 & 0.98 & $-$ \\ 
Gaia-T-ES9 & 1598160152636141568 & 87.67 & 49.03 & $4.54 \pm 0.42$ & 9.1 & 541.6 & 323.6 & -168.7 & 23898 & -3448 & 0.73 & $-$ \\ 
Gaia-T-ES10 & 2233912206910720000 & 88.96 & 13.49 & $3.59 \pm 0.23$ & 8.7 & 539.2 & 78.9 & -343.9 & 18526 & -419 & 0.84 & $-$ \\ 
Gaia-T-ES11 & 1042515801147259008 & 153.60 & 36.20 & $2.57 \pm 0.23$ & 10.0 & 523.4 & 403.2 & 73.9 & 16960 & -2075 & 0.72 & $-$ \\ 
Gaia-T-ES12 & 6385725872108796800 & 319.08 & -44.99 & $3.31 \pm 0.30$ & 6.8 & 547.0 & -32.1 & -11.5 & 13880 & -2307 & 0.68 & LMC? \\ 
Gaia-T-ES13 & 1552278116525348096 & 102.44 & 67.05 & $2.23 \pm 0.19$ & 8.5 & 530.4 & 490.3 & -83.6 & 13824 & -1401 & 0.70 & $-$ \\ 
Gaia-T-ES14 & 5195254636665583232 & 295.99 & -23.96 & $5.70 \pm 0.55$ & 7.7 & 536.3 & 506.3 & 191.8 & 13363 & 1055 & 0.63 & stream? \\ 
Gaia-T-ES15 & 5482348392671802624 & 269.29 & -28.85 & $7.54 \pm 0.70$ & 11.1 & 502.9 & 479.8 & 434.1 & 11715 & 1219 & 0.71 & stream? \\ 
Gaia-T-ES16 & 5190987741276442752 & 301.17 & -22.50 & $2.74 \pm 0.15$ & 7.1 & 536.9 & 513.5 & 171.1 & 8174 & -31 & 0.68 & $-$ \\ 
Gaia-T-ES17 & 5191438266165988352 & 299.83 & -21.25 & $5.84 \pm 0.48$ & 7.4 & 528.9 & 516.6 & 319.1 & 7255 & 350 & 0.57 & $-$ \\ 
Gaia-T-ES18 & 5637997011047611264 & 255.66 & 16.50 & $4.67 \pm 0.42$ & 10.2 & 502.6 & -236.2 & 252.5 & 6824 & -3386 & 0.63 & $-$ \\ 
Gaia-T-ES19 & 1765600930139450752 & 67.47 & -31.88 & $1.73 \pm 0.13$ & 7.6 & 523.6 & -376.3 & -271.8 & 4102 & 559 & 0.58 & $-$ \\ 
Gaia-T-ES20 & 330414789019026944 & 137.29 & -24.37 & $1.96 \pm 0.18$ & 9.4 & 504.4 & -150.9 & -120.6 & 3967 & -641 & 0.53 & $-$ \\ 
Gaia-T-ES21 & 5869501039771336192 & 307.46 & 1.96 & $3.07 \pm 0.30$ & 6.6 & 537.3 & 142.0 & 380.5 & 3374 & -1538 & 0.52 & LMC? \\ 
Gaia-T-ES22 & 1400950785006036224 & 75.16 & 52.41 & $6.21 \pm 0.58$ & 9.3 & 496.5 & 138.8 & 48.2 & 2495 & -1788 & 0.52 & LMC \\ 
Gaia-T-ES23 & 4747063907290066176 & 267.94 & -54.95 & $2.44 \pm 0.13$ & 8.4 & 508.9 & 276.2 & 25.3 & 2252 & -1440 & 0.55 & $-$ \\ 
Gaia-T-ES24 & 5373040581643937664 & 289.09 & 11.25 & $5.10 \pm 0.45$ & 8.0 & 511.5 & 501.1 & 333.8 & 177 & 417 & 0.49 & MWCenter? \\ 
Gaia-T-ES25 & 73753560659651584 & 152.87 & -45.72 & $1.29 \pm 0.07$ & 8.9 & 501.9 & 206.9 & -166.1 & -22 & 1050 & 0.47 & $-$ \\ 
Gaia-T-ES26 & 2260163008363761664 & 100.58 & 29.08 & $3.60 \pm 0.29$ & 9.3 & 494.5 & 352.5 & 14.6 & -928 & -1166 & 0.48 & $-$ \\ 
Gaia-T-ES27 & 1359836093873456768 & 72.19 & 37.90 & $3.76 \pm 0.32$ & 8.0 & 505.3 & -430.3 & -34.2 & -1564 & 1210 & 0.46 & $-$ \\ 
Gaia-T-ES28 & 2853089398265954432 & 108.61 & -36.13 & $1.45 \pm 0.13$ & 8.5 & 500.9 & 291.8 & -303.5 & -2531 & -3391 & 0.50 & $-$ \\ 
Gaia-T-ES29 & 5800686352131080704 & 316.14 & -23.51 & $3.81 \pm 0.24$ & 6.2 & 527.0 & 526.8 & 27.8 & -2570 & -73 & 0.45 & MWCenter \\ 
Gaia-T-ES30 & 4863753908114937728 & 229.68 & -52.12 & $3.66 \pm 0.35$ & 10.0 & 481.4 & -464.1 & -174.3 & -3203 & -1170 & 0.44 & $-$ \\ 
\enddata
\tablecomments{
The origin of the extreme velocity star is denoted as
MWCenter and LMC 
if it is consistent with coming from  
the Galactic center (center of the Milky Way) and the Large Magellanic Cloud, respectively. 
Gaia-T-ES14 and 15 have similar orbital properties, so they might be debris of the same system. 
}
\end{deluxetable*}

\begin{deluxetable}{l cc cc}
\tablecaption{Extreme velocity star sample with known metallicity
\label{table2}}
\tablewidth{0pt}
\tabletypesize{\scriptsize}
\tablehead{
\colhead{Short name} &
\colhead{[M/H]$_{\rm RAVE}$} &
\colhead{[M/H]$_{\rm LAMOST}$} &
\colhead{[Fe/H]$_{\rm HW2018}$} &
\colhead{Origin} 
}
\startdata
            &  dex & dex & dex \\ 
\hline 
Gaia-T-ES3  & $-1.37 \pm 0.18$ & $-1.139 \pm 0.194$ & $...$ & $-$  \\ 
Gaia-T-ES5  & $-1.80 \pm 0.14$ & $...$ & $...$ & $-$  \\ 
Gaia-T-ES8  & $-2.33 \pm 0.16$ & $...$ & $...$ & $-$  \\ 
Gaia-T-ES10 & $...$ & $...$ & $-1.72\pm0.16$ & $-$  \\ 
Gaia-T-ES13 & $...$ & $-0.957 \pm 0.214$ & $...$ & $-$  \\ 
Gaia-T-ES22 & $...$ & $-1.308 \pm 0.301$ & $...$ & LMC  \\ 
Gaia-T-ES27 & $...$ & $-1.425 \pm 0.228$ & $...$ & $-$  \\ 
\enddata
\tablecomments{
[M/H]$_{\rm RAVE}$ is obtained from RAVE Data Release 5. 
[M/H]$_{\rm LAMOST}$ is obtained from LAMOST Data Release 3. 
For Gaia-T-ES10, \cite{Hawkins2018} obtained 
spectroscopic metallicity [Fe/H]$_{\rm HW2018}$. 
}
\end{deluxetable}

\begin{figure}
\begin{center}
\includegraphics[angle=0,width=3.35in]{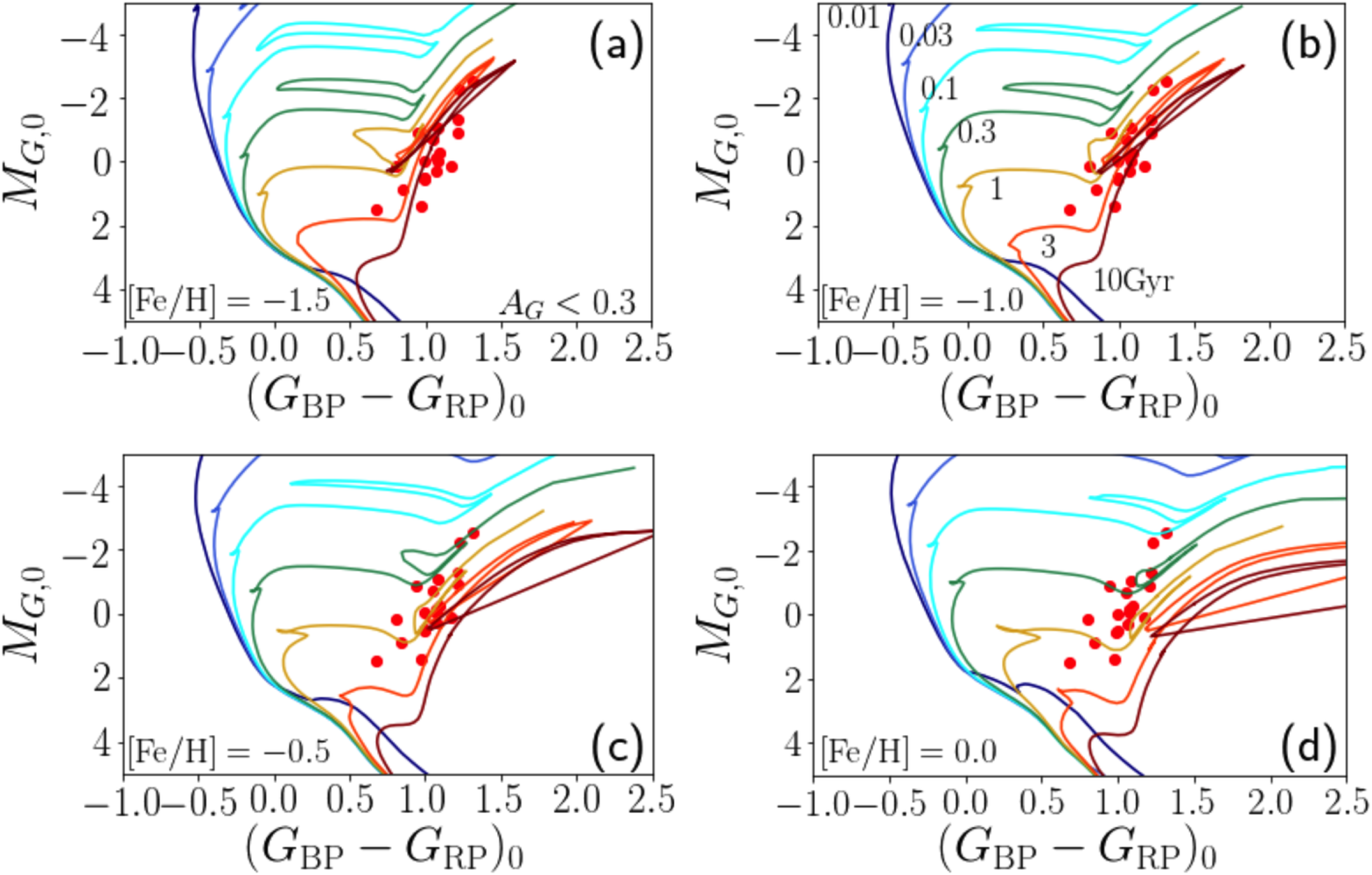} 
\end{center}
\caption{
\label{fig:CMD}
Dust-corrected color-magnitude diagram of a subset of our extreme velocity star sample 
with moderate dust extinction ($A_G<0.3$\,mag),  along with the PARSEC isochrone models with different metallicity and age. 
We can clearly see that most of our sample stars are evolved old stars and metal poor stars, suggesting that they do not originate in the stellar disk or 
the Galactic Center. 
}
\end{figure}

\section{Observed properties}
\label{sec:obsprop}
\subsection{Color-magnitude diagram}
\label{sec:CMD}

A color-magnitude diagram of a subset of 19 (out of 30) extreme velocity stars is shown in Figure \ref{fig:CMD}. 
Colors $(G_\mathrm{BP}-G_\mathrm{RP})$ and absolute magnitudes $M_G$
have been corrected, assuming the `combined' 3D dust model 
by \cite{Bovy2016mwdust}\footnote{
Available at \url{https://github.com/jobovy/mwdust}
}
and the point-estimate of the distance to our sample stars, $1/\varpi$. 
The detailed description for our dust correction is presented in 
Appendix \ref{sec:appendix_dust}. 
We have confirmed that 
using a different dust extinction model,
such as the 2D dust extinction model by \cite{Schlegel1998} 
recalibrated following \cite{Schlafly2011}, 
does not alter our main conclusion.\footnote{
We do not use the extinction values provided by {\it Gaia}, 
as they seem to overcorrect the colors and magnitudes of lower-luminosity giants.}
For stars with derived $A_G>0.3$\,mag, we are concerned that 
the uncertainty in 
$A_G$ might complicate the interpretation of our color-magnitude diagram, and so we discard these stars from Figure \ref{fig:CMD}, focusing instead on the 19 stars (shown in red) that happen to lie on lines of sight with less problematic extinction estimates\footnote{However, the 11 stars excluded from the plot do not differ significantly from the 19 stars shown.}. 
Figure \ref{fig:CMD} shows these extreme velocity stars superimposed on PARSEC isochrones 
\citep{Bressan2012}  
for stellar populations with age 0.01--10\,Gyr, and four different metallicities from 1/30 solar to solar metallicity.  
The isochrones suggest that 
the colors of the red giant branch and red clump stars 
are redder for higher metallicity, 
and thus the relatively blue colors of our sample stars 
are best explained if most of the extreme velocity stars 
are relatively old and metal poor with $-1.5<$~[Fe/H]~$< -1.0$ (panels (a) and (b)). 
However, we do not rule out the possibility that 
a minor fraction of stars in our sample might be metal-rich and young.

There is additional evidence supporting our claim that the stars in our sample are mostly old and metal-poor giants. 
According to the {\it Gaia}-DR2 catalog, Gaia-T-ES14 is classified as an RR Lyrae star (RRab star), which suggests that this star is old ($\sim 10 \Gyr$ old).
Also, we note that spectroscopic metallicity from RAVE Data Release 5 \citep{Kunder2017} and/or LAMOST Data Release 3 \citep{Cui2012,Zhao2012} is available for six stars in our sample,\footnote{
We used {\tt gaia\_tools} (\url{https://github.com/jobovy/gaia_tools}) to crossmatch LAMOST data with {\it Gaia} data. 
}  
and another star (Gaia-T-ES10) was recently observed by \cite{Hawkins2018}  
(see Table \ref{table2}). 
These seven stars show low metallicity of [M/H]$\lesssim -0.9$, which reinforces our argument that most of our sample stars are old and metal-poor.  
In addition, we note that the stellar radii reported in {\it Gaia}-DR2 archive 
for our 30 sample stars are larger than $3.8 R_\odot$, 
and those for 21 stars are larger than $10 R_\odot$. 
Although these stellar radii are not fully reliable 
(because they assume zero dust extinction; see \citealt{Andrae2018}), 
the reported large stellar radii of these stars support  
the idea that our sample is dominated by giants.

Our sample stars are clearly very different from previously known OB-type hypervelocity star candidates \citep{Brown2015ARAA} which are young and massive. 
We caution that our sample is restricted to apparently bright stars  
because we require that $\vlos$ is measured by {\it Gaia}. 
Therefore, our sample is highly biased in favor of intrinsically bright objects and is likely to be the tip of the iceberg in terms of the absolute magnitudes. 
However, because no cuts were made on the basis of metallicity, the fact that 
all of the stars with known metallicity are metal-poor 
and the other stars are consistent with being 
metal-poor and old, despite the fact that they are located in the solar neighborhood, is striking evidence that most of our sample stars are not a disk or Galactic Center population.

\begin{figure*}
\begin{center}
\includegraphics[angle=0,width=6.6in]{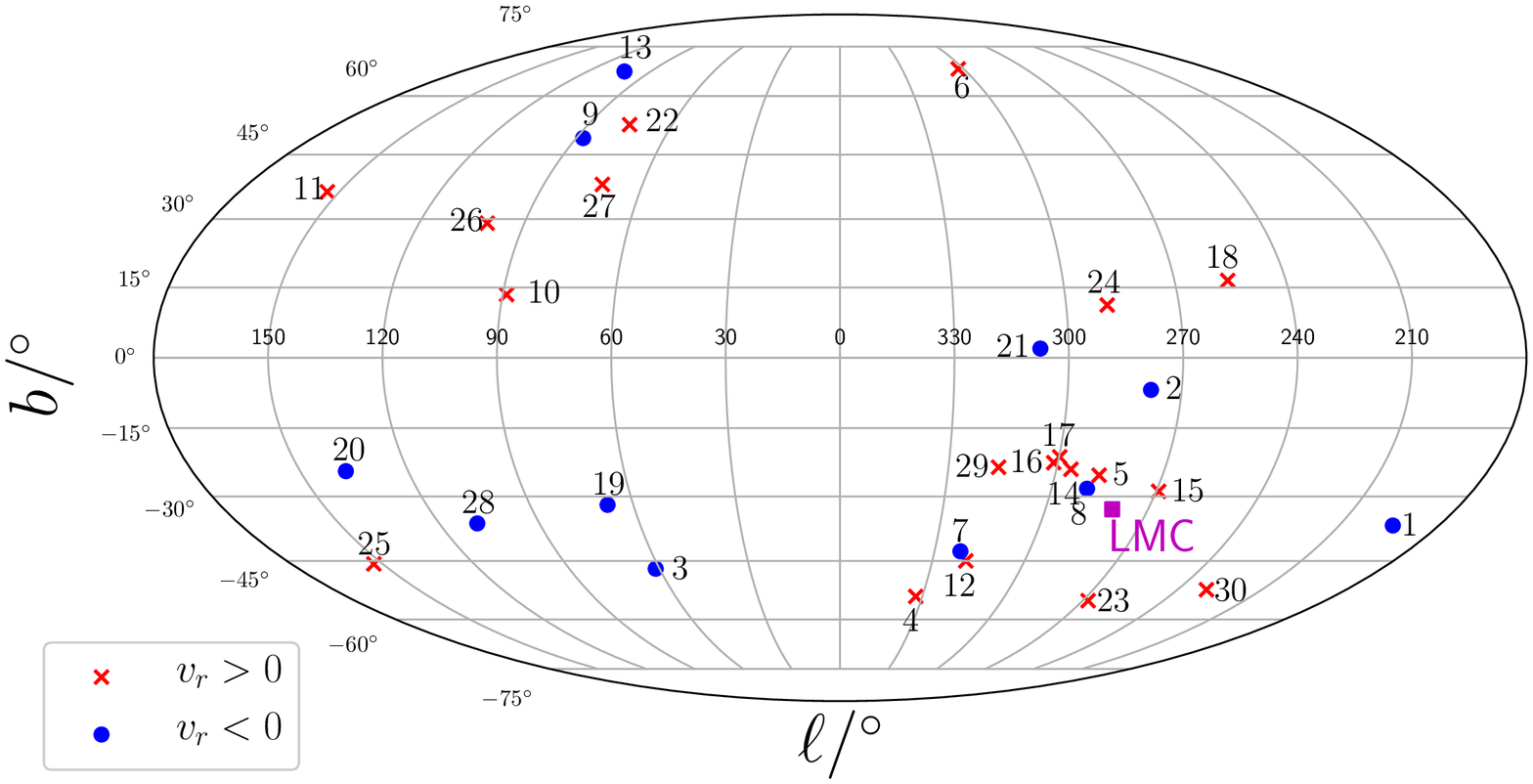} 
\end{center}
\caption{
\label{fig:sky}
The distribution of Galactic longitude and latitude of our extreme velocity star sample. 
Those stars with positive and negative Galactocentric radial velocity $v_r$ are marked with a red cross or a blue dot, respectively. 
The clumped region of our sample (with heliocentric distance $d_{\rm helio}<8 \kpc$) at around $(\ell, b) = (300^\circ, -30^\circ)$ is close to the projected location of the LMC. (Note, however, that the LMC is $d_{\rm helio}^{\rm LMC} \simeq 50 \kpc$ away.) 
Our orbital analysis suggests that Gaia-T-ES22 (and possibly Gaia-T-ES12 and 21) may have been ejected from the LMC (see also Figure \ref{fig:LMCorbit}). 
In addition, the orbit of Gaia-T-ES29 (and possibly Gaia-T-ES24) is consistent with originating from the center of the MW. 
}
\end{figure*}

\subsection{Distribution across the sky}
\label{sec:skyDistribution}

Figure \ref{fig:sky} shows the 30 extreme velocity stars in a Mollweide equal-area projection. It is clear that distribution of this sample on the sky is highly inhomogeneous. 
This inhomogeneous distribution is mainly due to our parallax precision cut. 
Because {\it Gaia} preferentially scans
regions with high ecliptic latitude\footnote{
See the 2D  {\it Gaia} Nominal Scanning Law available at \url{https://www.gaia.ac.uk/science/parallax/scan} \label{footnote:Gaia_scan}} 
(or the region roughly defined by $0 < \ell/^\circ < 180$ and $b>0^\circ$ 
as well as $180 < \ell/^\circ < 360$ and $b<0^\circ$), 
the typical quality of the astrometric solution is better for these regions, 
resulting in larger (a factor of $\sim2$) volume accessible with our parallax precision cut (see Appendix~\ref{sec:appendix_selection} for more detail). 
Even though the interpretation of the distribution of our sample stars on the sky is complicated due to this selection effect, it is remarkable that our sample covers the entire area on the sky, unlike the blue hypervelocity star candidates, which are mostly limited to the Northern sky (see Figure~7 of \citealt{Brown2014}).

Although \cite{Boubert2017} predicted that HVSs ejected from the LMC should produce a clustered distribution of stars on the sky (for stars located at heliocentric distances of $\sim 50 \kpc$), we caution that the apparent clustering of our extreme velocity sample near the LMC direction could be a mere coincidence. 
It is also clear from the radial velocities of this sample (indicated by the colored symbols in Fig.~\ref{fig:sky}) that most of the stars located near the LMC have radial velocities that indicate that they are moving radially outward---not inward, as would be expected if they  been coming from the LMC. 
In Section \ref{sec:orbits}, we  analyze the orbits of our sample stars to reinforce this argument.

\begin{figure}
\begin{center}
\includegraphics[angle=0,width=3.3in]{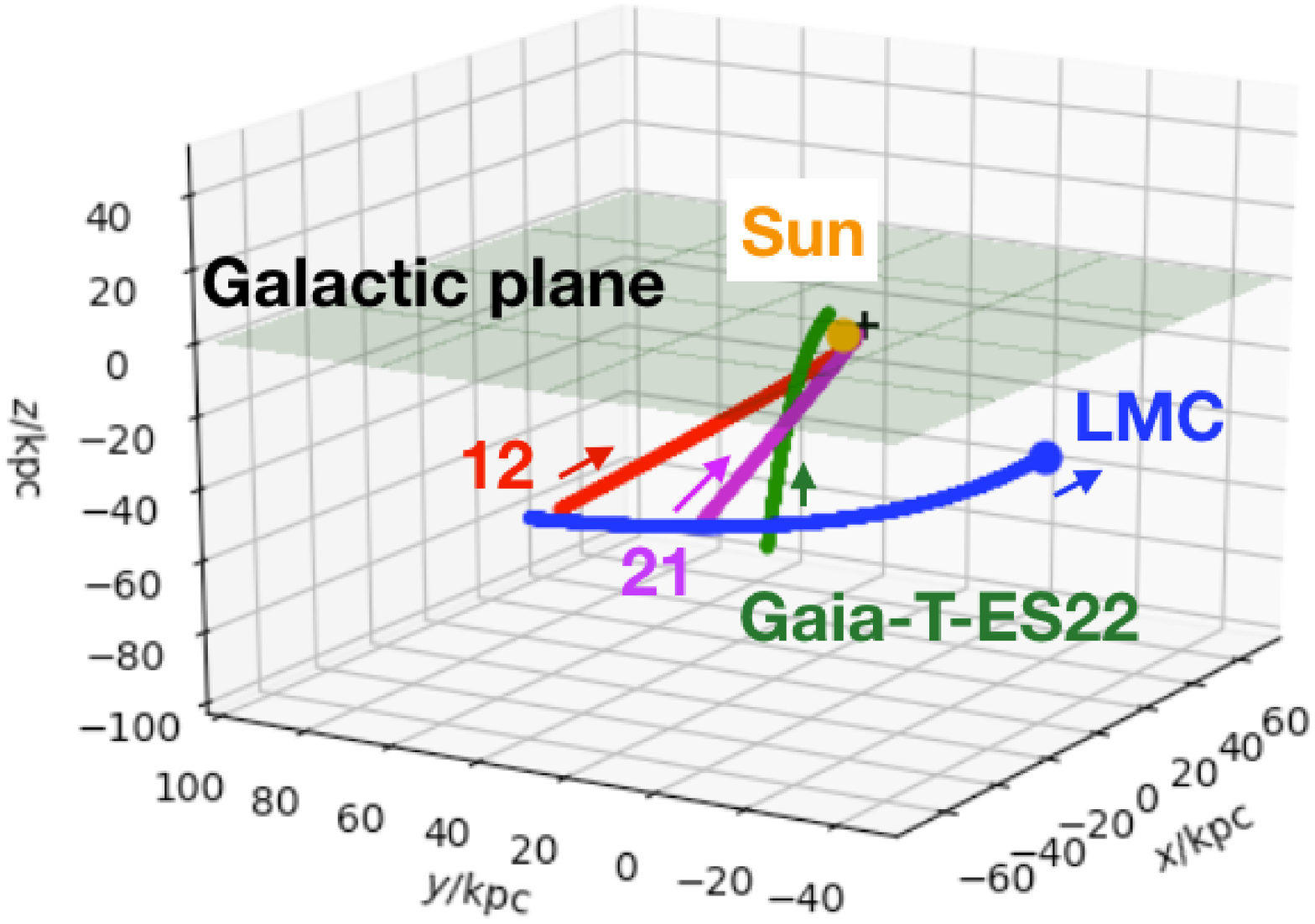} 
\end{center}
\caption{
\label{fig:LMCorbit}
The reconstructed orbits of the LMC and three of our sample stars (Gaia-T-ES12,21,22) with 
appreciable probability of having been ejected from the LMC. 
Here, the orbit of the LMC is integrated backward for $500 \Myr$, while the orbits of three stars are integrated backward until the time of closest approach to the LMC (which corresponds to the ejection epoch from the LMC). 
The current locations of the Sun and the LMC are marked with 
large orange and blue symbols, respectively. 
The shaded plane represents the Galactic disk plane, 
and the black plus sign (just behind the Sun) shows the location of the Galactic Center. 
}
\end{figure}

\subsection{Orbit analysis}
\label{sec:orbits}

In order to infer the origin of our sample stars, we compute orbits for all 30 stars in our sample, assuming the {\tt MWPotential2014} potential. 
Using the 6D 
observed quantities, along with their errors and correlations, 
we generate 1000 possible initial conditions for each star and evolve each one backward in time in the assumed MW potential, thus generating 1000 possible orbits for each star. 

Similarly, we compute the orbit of the LMC (integrated back in time) in the same Galactic potential by assuming that the LMC is a test particle. 
We generate 1000 orbits by taking into account the current 
observed coordinate   
of the LMC; 
we assume that the line-of-sight velocity of the LMC is $\vlos = 262.2 \pm 3.4 \kms$ \citep{vanderMarel2002}, 
its proper motion is $(\mualpha, \mudelta) = (1.850 \pm 0.03, 0.234 \pm 0.03) \masyr$ \citep{Gaia-Helmi2018}, 
and its distance modulus is $18.50\pm0.10$ \citep{Freedman2001}. The orbits are then used to infer the probability that a given star was ejected from the Galactic Center, MW disk, or the LMC.

First, we find that 18 stars (Gaia-T-ES1-4, 9, 13-17, 20-22, 24, 26-29) have crossed the stellar disk of the MW at $r< 30 \kpc$ in the past. 
One of these stars, Gaia-T-ES29, has a probability $P(r< 0.25 \kpc)= 0.55$ of crossing within $0.25 \kpc$ from the Galactic Center, and the flight time 
(measured along the orbit)  
from the Galactic Center to the current location is typically $\sim 9 \Myr$. 
Another star, Gaia-T-ES24, has a probability of $P(r< 0.25 \kpc)= 0.08$ and typical flight time of $\sim 2 \Gyr$. 
(We note that Gaia-T-ES24 is a marginally bound star, 
and this star comes close to the Galactic Center 
only if we integrate the orbit backward in time for as long as $\sim 2\Gyr$.) 
Thus, these two stars (or at least Gaia-T-ES29) are consistent with having been ejected from the Galactic Center.\footnote{
We note that \cite{Marchetti2018} classified Gaia-T-ES7 as consistent with having been ejected from the Galactic Center. However, in our potential model, we found that the probability $P(r< 0.25 \kpc)$ is consistent with zero for this star, and $P(r< 1 \kpc)$ is as small as $0.002$. This result demonstrates that adopting a slightly different potential model can affect the inference about the origin of our sample stars.
} 
For the other 16 stars, we examined each disc crossing velocity, but all the disc crossing velocities are too large to be consistent with the ordinary disk ejection mechanisms (such as SN ejection or dynamical ejection in star clusters, which can eject giant stars with a velocity of  at most $100 \kms$ in the frame of the disk streaming motion).
While other mechanisms \citep{gualandris_portegies-zwart_07,Gvaramadze09} that operate in young massive star clusters (interaction with an intermediate-mass black hole (IMBH) or a supermassive star both formed from runaway mergers of massive young stars) could eject giant stars with $v_{\rm ej} > 600 \kms$, it is unlikely that they were ejected recently from such a young cluster, given that our stars are old 
and metal poor. Furthermore, 
if these had been ejected from the stellar disk, the same mechanism should have also ejected younger and more metal-rich stars, which are not found in our kinematically selected sample.

Second, we find that three stars (Gaia-T-ES12, 21, 22) have orbits that have finite probability of having been ejected from the LMC. 
The most likely candidate star from the LMC is Gaia-T-ES22, 
which has a probability $P(d_\mathrm{LMC}<5 \kpc)= 0.27$ of passing within $5 \kpc$ of the LMC at around $200 \Myr$ ago. 
At the epoch of closest approach to the LMC, the relative velocity of 
Gaia-T-ES22 with respect to the LMC's center of mass is $\sim 200 \kms$. 
Taking into account the fact that the stellar disk of the LMC rotates at $\sim (80$-$90) \kms$ with respect to the LMC's center of mass \citep{Gaia-Helmi2018,Vasiliev2018LMC}, the ejection velocity of $\sim 200 \kms$  may not require ejection by a (hypothetical) massive black hole at the center of the LMC. 
Also, Gaia-T-ES12 and 21 have respective probabilities of $P(d_\mathrm{LMC}<15 \kpc)= 0.02$ and $0.06$ of passing within $15 \kpc$ of the LMC, and their closest approaches are about $500$ and $300 \Myr$ ago, respectively. 
Figure \ref{fig:LMCorbit} illustrates the reconstructed orbits of these three stars and of the LMC. 
We note that other authors have also identified escaping stars from the LMC from different catalogs. For example, \cite{Lennon2017} found a supergiant star whose velocity is consistent with originating from the LMC, and \cite{Erkal2018} found that a hypervelocity star candidate known as HVS3 \citep{Brown2015ARAA} has a high probability of having been ejected from the LMC.

It is important to point out that none of the orbits of the other stars located in the region of high clustering around the LMC (such as Gaia-T-ES5, 8, 14-17) have orbits that came close to the LMC in the past.  
In contrast, Gaia-T-ES22, which has the highest probability of having been ejected from the  LMC in our sample, is located on a part of the sky where \cite{Boubert2016} predict the lowest density of HVSs from the LMC. 
We also note that Gaia-T-ES14 and 15 have similar orbital energies, angular momenta, positions, and velocities. Thus, these two stars might belong to an unknown stellar stream that happens to be located near the line of sight to the LMC.

A more sophisticated model that includes the gravitational potential of the LMC 
(see \citealt{Erkal2018} for an example of such a model)
is required before a definitive statement can be made about where in the LMC these stars were ejected from. 
However, our tentative result that at least one star has an orbit consistent with the LMC merits further investigation, given that the orbits may depend on the assumed model for the MW potential.

\section{Discussion}
\label{sec:discussion}

\subsection{Mechanisms for accelerating stars with extreme velocities}
\label{sec:mechanisms}

We now discuss a few possible mechanisms for accelerating stars to extremely high velocities, and evaluate the likelihood that the stars in our sample were accelerated this way. 
Given that only one or two of the stars in our sample are consistent with having been ejected from the Galactic Center, we do not discuss the Hills mechanism, which is considered to be responsible for ejection of blue HVSs.  

\paragraph{Ejection of the stellar binary companion of a Type Ia Supernova:} 

Type Ia supernovae (SNe~Ia) are thought to arise from the thermonuclear ignition and burning of a C/O white dwarf in a binary system. The event could be triggered by accretion from a main sequence or giant star companion (single-degenerate scenario) or from another white dwarf (double-degenerate scenario). Liberation of giant companions of Type II supernovae (SNe II) was proposed to explain runaway OB stars \citep{Blaauw61}. A similar process can also result in the ejection of companions of SN Ia progenitors. \citet{Shen2018} reported the discovery of three hypervelocity white dwarfs in the {\it Gaia}-DR2 sample that  they propose are the liberated companions of double-degenerate SN Ia. The maximum ejection velocity of the companion star in such a scenario depends on the minimum orbital radius $r_{\rm min}$ (because one can assume that the SN~Ia progenitor has a mass of $\leq 1.4$\msun), with $r_{\rm min}$ limited by the radius of the companion star. The ejection velocity of a white dwarf in the double-degenerate scenario can be as high as several $\sim 1000 \kms$ \citep{Shen2018}. A stellar companion of solar radius or smaller can achieve an ejection velocity of $\sim 600 \kms$. In contrast, ejection velocities of giant stars (radii $\sim 10$--$30R_\odot$) are expected to be significantly smaller ($\sim 60$--$100\kms$). 

While the precise radii of the extreme velocity stars in our sample are uncertain, their locations on the color-magnitude diagram indicate that they should be greater than 10$R_\odot$. Their observed space velocities ($\gtrsim 480$\kms) are clearly too large for them to have been accelerated to their observed velocities following a recent ejection from the MW disk.  

However, because the stars in our sample have ages and metallicities that are consistent with the populations of globular clusters, where the probability of forming stellar binaries is high, it is possible that these stars were once main sequence companions of single degenerate SN~Ia that detonated inside globular clusters. Because globular clusters orbit the halo with space velocities of $\sim 300$-$400 \kms$ \citep{Gaia-Helmi2018}, the stellar companion of such a SN~Ia 
that is ejected when the globular cluster is near its apocenter could attain a velocity of $\sim 500 \kms$ by the time it passes through the solar neighborhood (which is near the pericenter of the orbit).

The ages and metallicities of stars with possible LMC origin are similar to the rest of the sample and also consistent with the ages and metallicities of LMC globular clusters \citep{Beaulieu99}. It is therefore plausible that the extreme velocity stars in our sample were once members of stellar binaries in globular clusters, either in the MW or in the LMC.

According to LAMOST Data Release 3, 
Gaia-T-ES22, whose orbit is consistent with originating from LMC, 
has a low metallicity of [M/H]=$-1.308\pm0.301$. 
This metallicity corresponds to the low-metallicity tail of metallicity distribution of the LMC's inner stellar disk \citep{Pompeia2008,Olsen2011}, 
and consistent with the metallicities of some globular clusters near the central region of the LMC such as NGC 1898 \citep{Olsen1998, Johnson2006}, NGC 1928 \citep{Mackey2004}, and NGC 2019 \citep{Olsen1998, Grocholski2006, Johnson2006}. 
In order to better understand the origin of this star, 
we must obtain detailed chemical information on Gaia-T-ES22 and compare it to  other stars in the LMC.

\paragraph{Interaction of stellar binaries with an IMBH:}

Super massive stars of $800$-$3000\msun$ have been proposed to form as a result of runaway mergers of individual stars \citep{Portegies_Zwart04} or via three-body encounters of massive stars with stellar binaries \citep{gurkan_etal_06} in young dense star clusters. Such massive stars can ultimately collapse to form an IMBH within about $10 \Myr$. The presence of such an IMBH in a dense star cluster could result in frequent strong encounters with stellar binaries. 
Assuming that young dense star clusters can contain IMBHs of mass between $10^2 \msun$ and $10^4 \msun$, \citet{gualandris_portegies-zwart_07} used a large suite of simulations to show that it was possible to achieve ejection velocities of $v_{\rm ej} >500 \kms$ for $M_{\rm IMBH} >10^3\msun$ with the Hills mechanism. 
However, given that the extreme velocity stars in our population are old and metal-poor, it seems unlikely that they were ejected by IMBHs in young star clusters in either the MW disk or the LMC. 

It has also been proposed that IMBHs could grow via binary star interactions with $\sim 50~\msun$ black holes in globular clusters \citep{miller_hamilton_02}. 
Dynamical modeling of the kinematics of stars at the centers of globular clusters has also been used to argue for the existence of $10^3$--$10^4\msun$ IMBHs in globular clusters like M15 (in the MW) and G1 (in M31) \citep[e.g.,][]{gressen_etal_02, gebhardt_etal_05,lutzgendorf_etal13}. 
However, these results are controversial because the kinematical data can generally also be explained by a dense concentration of stars instead of an IMBH \citep{vandenbosch_06}, and stringent limits on the continuum radio flux from possible IMBHs at the centers of globular clusters \citep{Strader_Chomiuk2012} imply that  their accretion rates are extremely low, if they exist. If an IMBH resides at the center of a globular cluster, it quite could easily eject stars via the Hills mechanism.

\paragraph{Tidal stream debris from satellites:}
Many of the dwarf spheroidal satellites of the MW, as well as the outskirts of the LMC, have relatively old and metal-poor stellar populations. When a satellite is disrupted, some of the stars become bound to the MW, and some become unbound and eventually escape. 
It is therefore plausible that some of the extreme velocity stars could be associated with tidal debris from accreted satellites that traveled close to the Galactic Center \citep{Abadi2009,Teyssier_etal_09}, or accreted material from the LMC. 
Detailed stellar abundances of the stars in our sample may help to make a more definitive statement regarding their origin. 

\subsection{Arguments for a higher escape velocity}
\label{discussion_esc}

In order to explore the implications of the observed space density of extreme velocity stars, we compare possible production rates of ejected stars from globular clusters (via SN Ia or IMBH-binary star ejections) with rates needed to replenish the extreme velocity star population if they are all escaping. 
Assuming that $\sim 25$ stars 
not originating from the LMC were ejected from globular clusters in the MW, and based on their observed volume density (within a sphere of radius $8 \kpc$), 
we estimate that there should be $\sim 10^3$ similarly old, metal-poor giant stars within 30~kpc (the radius containing most of the MW globular clusters). Correcting for the rarity of giant stars (using PARSEC isochrones), we estimate a total of $\sim 10^6$ extreme velocity stars within 30~kpc. 
If unbound, these stars would escape from this region in $\sim 10^8$~yr and would require an SN Ia ejection rate of  $\sim 10^{-2}$~yr$^{-1}$. 
Based on the work of \cite{Voss_Nelemans_12}, we optimistically estimate the SN~Ia rate in the globular cluster population to be at most $\sim 10^{-4}$~yr$^{-1}$, a factor of 100 too small to account for the observed number of extreme velocity stars.

\citet{gualandris_portegies-zwart_07} find that interactions of stellar binaries with a $10^3\msun$-IMBH occurring with an impact parameter of $<1 \; \mathrm{au}$ can result in the capture of one star and the ejection of the other with a probability of 0.5. If IMBHs do exist in globular clusters, assuming typical values for the central stellar velocity dispersion ($\sim 10$\kms) and central stellar number density ($10^3 \pc^{-3}$), it is straightforward to estimate that the rate at which  stars can be ejected by the IMBH is no more than
$\sim 10^{-8}$~yr$^{-1}$ per globular cluster, which is orders of magnitude too low to account for the observed population of extreme velocity stars.

If the observed extreme velocity stars are indeed unbound, both mechanisms fall  factors of 100 or more short of being able to produce extreme velocity stars at the rate necessary to compensate for their escape from the MW.

In fact, the simplest  explanation for the observed population is that, regardless of how they are accelerated to these velocities, the observed stars are in fact bound to the MW. 
In Figure~\ref{fig:r_Vtot}, the blue dashed line shows the escape velocity curve for the {\tt MWPotential2014} potential, while the brown and magenta dashed lines show the escape velocity curves for two higher mass halos (with the same baryonic mass distribution and almost identical rotation curves within $r<8 \kpc$). 
It is clear that a MW that is $\sim 2$ times more massive than the {\tt MWPotential2014} model is massive enough to bind the stars in our sample. 
Based on this figure, we estimate a local escape speed of $v_{\rm esc} \sim 600 \kms$. 
We also tentatively estimate the virial mass to be $M_{200} \sim 1.4\times 10^{12}\msun$, although the estimate for $M_{200}$ clearly needs more sophisticated analysis and should be compared with the dark halo mass derived from other methods (e.g., \citealt{Gnedin2010, Penarrubia2016}).  Our value is in the middle of the range of recent values and consistent with two very recent estimates obtained using proper motions of halo globular clusters and satellites obtained with {\it Gaia} DR2 \citep{posti_helmi2018,Watkins2018}.
We also note that the number of our sample stars with positive and negative $v_r$ are more or less comparable (see Table \ref{table1}), which is reasonable if these stars are a bound population.

Our rough estimate of the local escape velocity is somewhat larger than previous measurements based on sophisticated modeling of local populations of stars. 
For example, \cite{Piffl2014} used RAVE data to derive\footnote{
We note that \cite{PifflBinney2014} reports $v_{\rm esc} = 613 \kms$, but they argue in their footnote 4 that the value of $v_{\rm esc}$ in \cite{Piffl2014} is more robust. 
} 
$v_{\rm esc} = 533^{+54}_{-41} \kms$, 
while 
\cite{Williams2017} used SDSS data to derive $v_{\rm esc} = 521^{+46}_{-30} \kms$. 
After the submission of this paper, 
\cite{Monari2018} estimated the local escape velocity of 
$v_{\rm esc} = 580 \pm 63 \kms$ using {\it Gaia} DR2, 
which is consistent with our estimate. 
They also estimated a virial mass of 
$M_{200} = 1.28^{+0.68}_{-0.50} \times 10^{12} \msun$, 
which is consistent with our value.

\section{Conclusions}
\label{conclusions}

We have discovered 30 new extreme velocity stars in the {\it Gaia}-DR2 archive.  
Our sample size is comparable to the number of known blue HVSs in the distant halo. 
A comparison of the dust-corrected color-magnitude diagram for this sample with the PARSEC isochrones indicates that, unlike previously discovered blue HVSs, these stars are old, metal-poor, and are most similar to the stellar populations in globular clusters 
or in the stellar halo. Using 6D phase space coordinates from {\it Gaia}, we compute the orbits of all the stars in our sample and conclude that up to three of the stars are consistent with having been ejected from the LMC, one or two stars are consistent with having been ejected from the Galactic Center, and the rest are halo objects of currently undetermined origin. 
Because these stars are bright, detailed abundances can yield more evidence on their origin.

While these stars have space velocities implying that they are unbound in the {\tt MWPotential2014} potential \citep{Bovy2015}, they would be bound if the local escape velocity is $\sim 600 \kms$ 
(which is higher than 
pre-{\it Gaia} 
estimates by $\sim 13 \%$, 
but consistent with the estimate with {\it Gaia} DR2 by \citealt{Monari2018}). 
This might also imply that the dark matter mass of the MW is $M_{200} \sim 1.4 \times 10^{12} \msun$, which is $\sim 2$ times larger than that of {\tt MWPotential2014} but completely consistent with two recent estimates
obtained with kinematics of globular clusters from {\it Gaia} DR2 \citep{posti_helmi2018,Watkins2018}.

\acknowledgments

We thank the referee for helpful comments 
and for some ADQL queries. 
We thank members of the stellar halos group at the University of Michigan for stimulating discussion. 
We thank Oleg Gnedin for several useful discussions and suggestions. 
We thank Warren Brown, David Katz, Scott Kenyon, and Tommaso Marchetti for their useful comments on the manuscript. 
K.H. thanks Keith Hawkins, Andy Casey, and Vasily Belokurov for useful conversation. 
M.V. thanks Kayhan Gultekin for discussions on IMBH-binary interaction rates. 
M.V. and K.H. are supported by NASA-ATP award NNX15AK79G.
I.U.R. acknowledges support from
grants PHY~14-30152 (Physics Frontier Center/JINA-CEE), 
AST~16-13536, and AST~18-15403
awarded by the U.S.\ National Science Foundation (NSF).~
This research was started at the NYC {\it Gaia} DR2 Workshop at the Center for Computational Astrophysics of the Flatiron Institute in 2018 April. 
This work has made use of data from the European Space Agency (ESA)
mission {\it Gaia} 
(\url{http://www.cosmos.esa.int/gaia}), 
processed by the {\it Gaia} Data Processing and Analysis Consortium (DPAC,
\url{http://www.cosmos.esa.int/web/gaia/dpac/consortium}). 
Funding for the DPAC has been provided by national institutions, in particular
the institutions participating in the {\it Gaia} Multilateral Agreement.
Funding for RAVE has been provided by: the Australian Astronomical Observatory; the Leibniz-Institut fuer Astrophysik Potsdam (AIP); the Australian National University; the Australian Research Council; the French National Research Agency; the German Research Foundation (SPP 1177 and SFB 881); the European Research Council (ERC-StG 240271 Galactica); the Istituto Nazionale di Astrofisica at Padova; The Johns Hopkins University; the National Science Foundation of the USA (AST-0908326); the W. M. Keck foundation; the Macquarie University; the Netherlands Research School for Astronomy; the Natural Sciences and Engineering Research Council of Canada; the Slovenian Research Agency; the Swiss National Science Foundation; the Science \& Technology Facilities Council of the UK; Opticon; Strasbourg Observatory; and the Universities of Groningen, Heidelberg and Sydney. The RAVE web site is at \url{https://www.rave-survey.org}.
Guoshoujing Telescope (the Large Sky Area Multi-Object Fiber Spectroscopic Telescope LAMOST) is a National Major Scientific Project built by the Chinese Academy of Sciences. Funding for the project has been provided by the National Development and Reform Commission. LAMOST is operated and managed by the National Astronomical Observatories, Chinese Academy of Sciences.
This research has also made use of NASA's
Astrophysics Data System Bibliographic Services;
the arXiv pre-print server operated by Cornell University;
the SIMBAD and VizieR
databases hosted by the
Strasbourg Astronomical Data Center.


\software{
Agama \citep{Vasiliev2018},\;
{\tt gaia\_tools} (\url{https://github.com/jobovy/gaia_tools}), 
matplotlib \citep{hunter07}, 
mwdust \citep{Bovy2016mwdust},  
numpy \citep{vanderwalt11}, 
scipy \citep{jones01}} 

\bibliographystyle{aasjournal}
\bibliography{mybibtexfile}

\appendix

\section{Dust correction}
\label{sec:appendix_dust}

In this Appendix, we describe how we derive the 
dust-corrected colors $(G_\mathrm{BP} - G_\mathrm{RP})_0$ 
and magnitudes $G_0$ for our sample stars, 
based on the values of $A_V^\mathrm{model}$ and $A_I^\mathrm{model}$ 
provided by \cite{Bovy2016mwdust}, 
as functions of 3D position of the stars.

First, we use the relationships in Appendix A of \cite{Evans2018arXiv180409368E} 
to derive the colors $(G-V)$ and $(G-I)$:
\eq{
(G-V) 
&= a_0 
+  a_1 (G_\mathrm{BP} - G_\mathrm{RP}) 
+  a_2 (G_\mathrm{BP} - G_\mathrm{RP})^2 \\
(G-I) 
&= b_0 
+  b_1 (G_\mathrm{BP} - G_\mathrm{RP}) 
+  b_2 (G_\mathrm{BP} - G_\mathrm{RP})^2 ,
}
where $(a_0, a_1, a_2) = (-0.01760, -0.006860, -0.1732)$ 
and $(b_0, b_1, b_2) = (0.02085, 0.7419, -0.09631)$. 
Next, the dust-corrected $V_0$, $I_0$, and $(V-I)_0$ are given by 
\eq{
V_0 &= G - (G-V) - A_V^\mathrm{model} \\
I_0 &= G - (G-I) - A_I^\mathrm{model} \\
(V-I)_0 &= V_0 - I_0 . 
}
By using the relationships in Appendix A of \cite{Evans2018arXiv180409368E}, 
we find that the dust-corrected $G_0$ and $(G_\mathrm{BP} - G_\mathrm{RP})_0$ are given by 
\eq{
G_0 &= V_0 
+ c_0 
+ c_1 (V-I)_0
+ c_2 (V-I)_0^2 
+ c_3 (V-I)_0^3 , \\
(G_\mathrm{BP} - G_\mathrm{RP})_0 &= 
  d_0
+ d_1 (V-I)_0
+ d_2 (V-I)_0^2 ,
}
where $(c_0, c_1, c_2, c_3) = (-0.01746, 0.008092, -0.2810, 0.03655)$ 
and 
$(d_0, d_1, d_2) = (-0.04212, 1.286, -0.09494)$. 
The dust extinction values $A_G$ and $E(G_\mathrm{BP} - G_\mathrm{RP})$ 
are given by 
\eq{
A_G &= G - G_0 \\
E(G_\mathrm{BP} - G_\mathrm{RP}) &= (G_\mathrm{BP} - G_\mathrm{RP}) - (G_\mathrm{BP} - G_\mathrm{RP})_0 . 
}
We note that our dust correction results in $A_G \simeq 0.8 A_V$, 
which is consistent with the calculations by \cite{Jordi2010}.

\section{Selection effect in high parallax-precision sample}
\label{sec:appendix_selection}

As mentioned in Section \ref{sec:skyDistribution}, our sample may be affected by our cut of $\varpi/\delta\varpi>10$ (high signal-to-noise ratio of parallax). 
Here, we illustrate
this selection effect.

\begin{figure}
\begin{center}
\includegraphics[angle=0,width=3.3in]{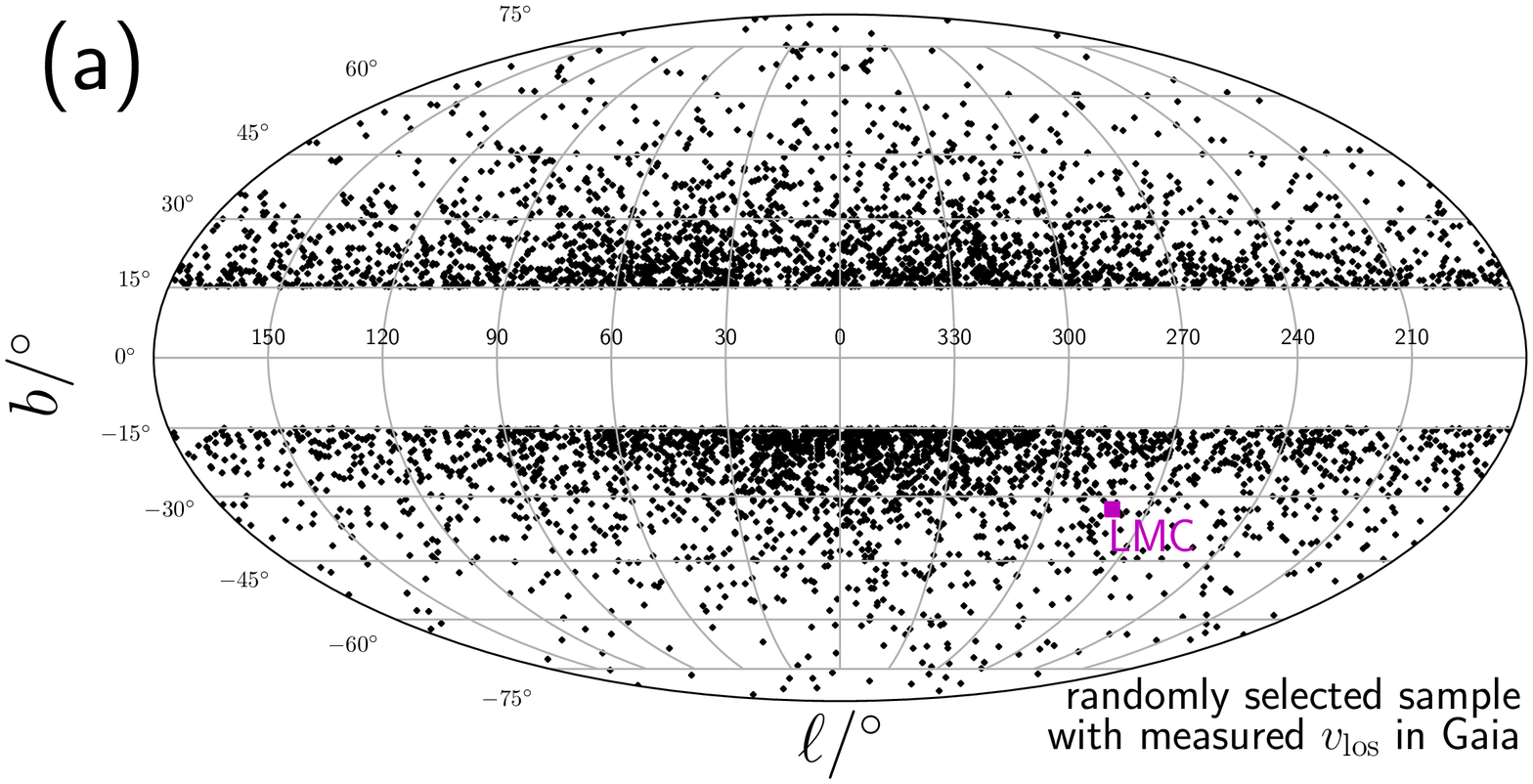} 
\includegraphics[angle=0,width=3.3in]{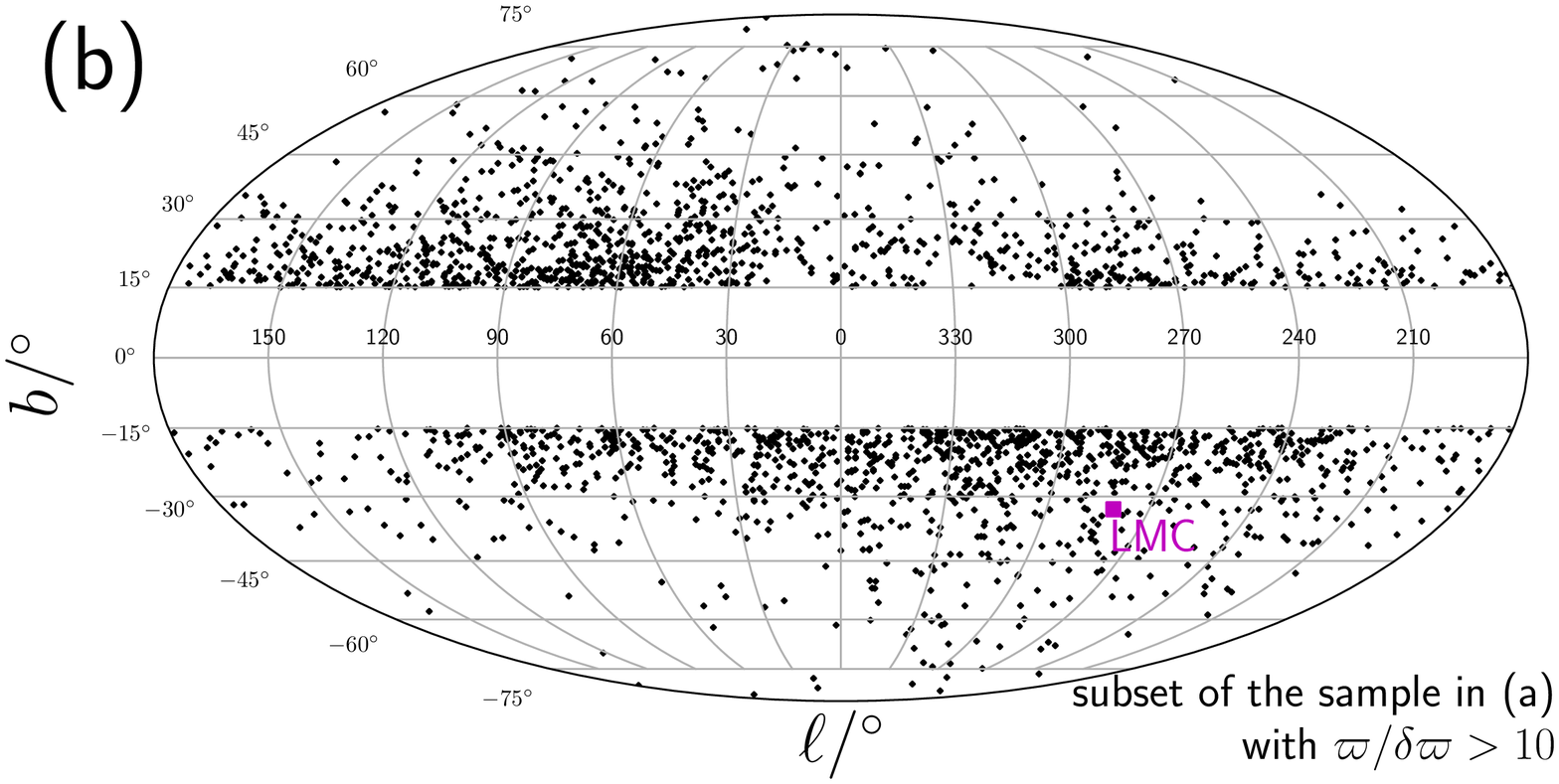} \\
\includegraphics[angle=0,width=3.3in]{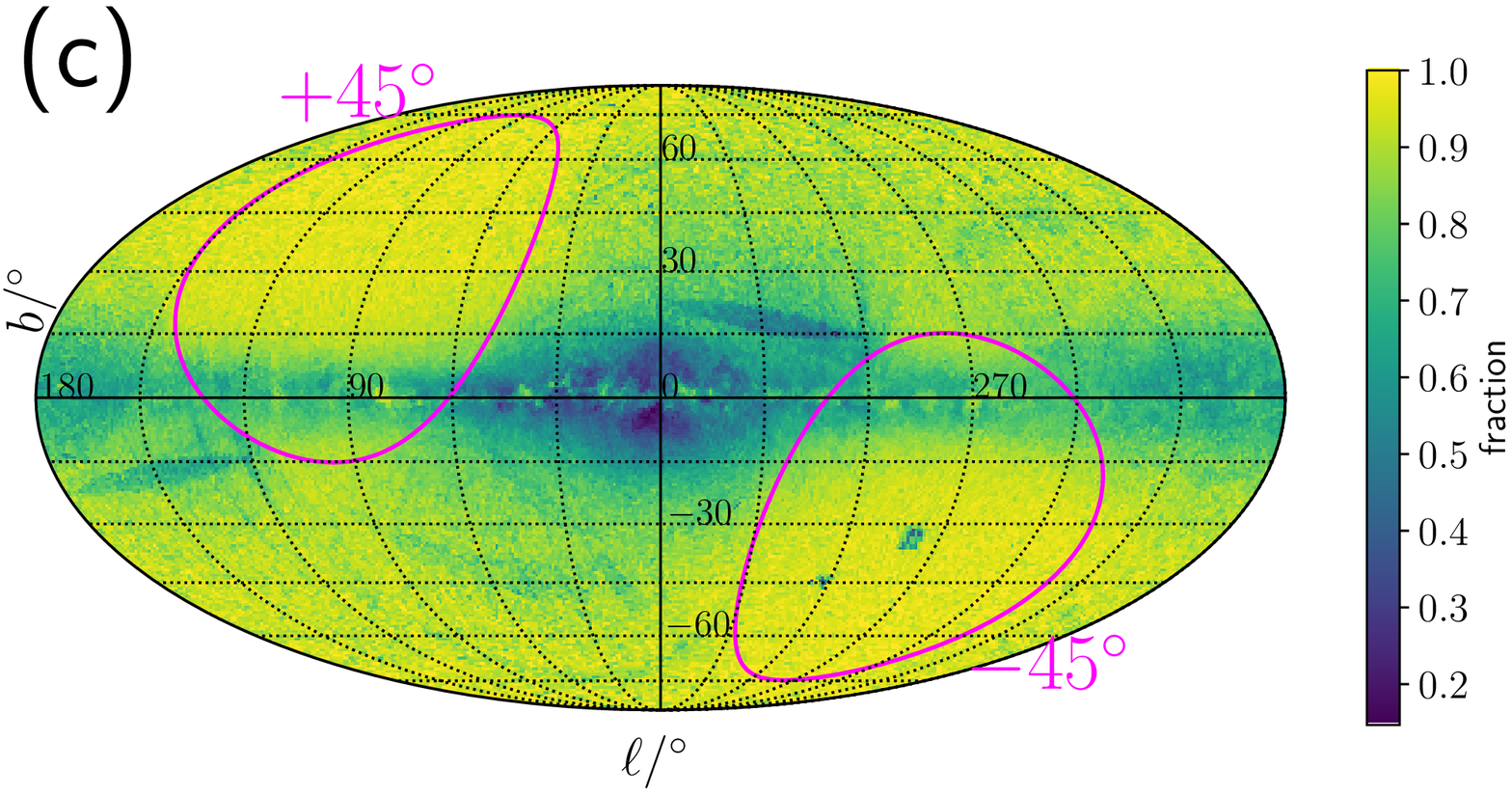}
\includegraphics[angle=0,width=3.3in]{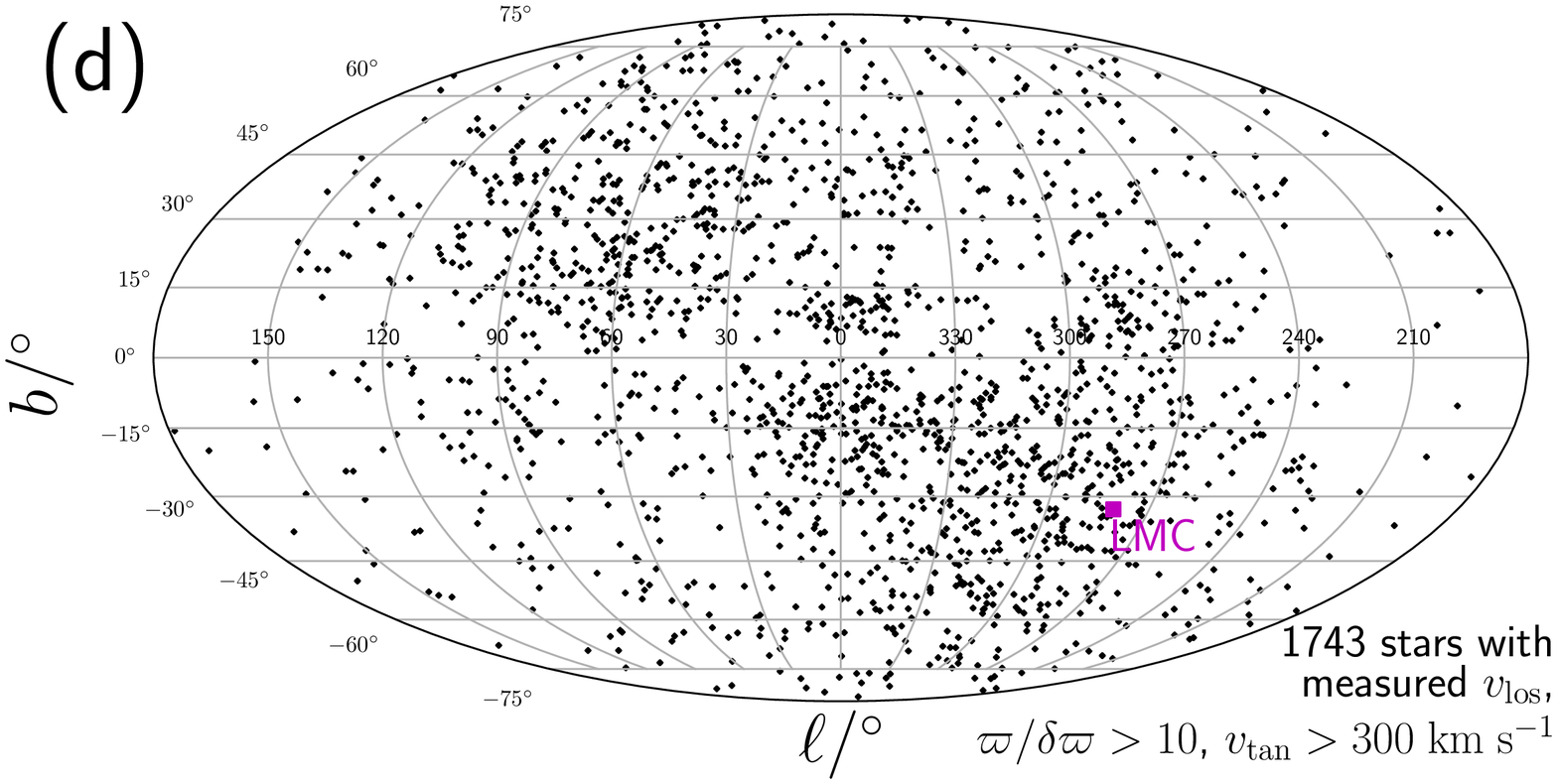}
\end{center}
\caption{
\label{fig:selectionEffect}
Illustration of the selection effect in our analysis. 
(a) Randomly selected stars at $|b|>15^\circ$ with measured $\vlos$ in {\it Gaia} DR2 and with $\varpi<0.5 \mas$. 
(b) A subset of stars on panel (a) with good parallax ($\varpi/\delta\varpi>10$). 
(c) 
The fraction of high-precision parallax stars ($\varpi/\delta\varpi>10$) 
in the {\it Gaia}-DR2 sample with $\vlos$ measurement. 
The magenta curves denote the ecliptic latitude contours of 
$+45^\circ$ and $-45^\circ$. 
We see that regions with high ecliptic latitudes 
show a higher fraction of stars with good parallax
due to more frequent observation by {\it Gaia}. 
(d) Our initial sample of 1743 stars with measured $\vlos$, good parallax ($\varpi/\delta\varpi>10$), and large tangential velocity $v_{\rm tan}>300 \kms$ (see Section \ref{sec:sample}). 
As a reference, we plot the location of the LMC in panels (a),(b), and (d). 
}
\end{figure}

Figure \ref{fig:selectionEffect}(a) shows the distribution of randomly selected stars ($N=5089$) in {\it Gaia} DR2 
with measured line-of-sight velocities ($\vlos$) that have $|b|>15^\circ$ and $\varpi<0.5 \mas$. (We note that this sample is completely independent from the kinematically selected sample in the main text.)  
We see that the distribution of the stars is symmetric around both $\ell=0^\circ$ and $b=0^\circ$, reflecting the symmetry of the stellar disk.

Figure \ref{fig:selectionEffect}(b) shows the distribution of a subset of stars ($N=2271$) in panel (a) that have high-precision parallax, $\varpi/\delta\varpi>10$. 
As we can clearly see, our cut in $\varpi/\delta\varpi$ introduces an asymmetric distribution of stars across the sky. 
At $180<\ell/^\circ<360$ (right-hand side of the panel), 
we see more stars with high-precision parallax at $b<0^\circ$ (near the LMC direction), 
while an opposite asymmetry is seen at $0<\ell/^\circ<180$.

These results can also be visualized in a slightly different manner. 
Figure \ref{fig:selectionEffect}(c) 
shows the fraction of stars with good parallax measurement 
($\varpi/\delta\varpi>10$)
in the {\it Gaia}-DR2 sample stars with measured $\vlos$. 
In evaluating this fraction, we derived for each line of sight:  
(i) the number density of {\it Gaia} sample stars with measured $\vlos$; and 
(ii) that with measured $\vlos$ and good parallax ($\varpi/\delta\varpi > 10$). 
The ratio of the latter density to the former 
is shown 
as a function of $(\ell,b)$. 
The magenta curves correspond to the ecliptic latitudes of $+45^\circ$ and $-45^\circ$. 
We see that our selection criterion of high-precision parallax disfavors 
high-density regions (e.g., near the Galactic plane or the LMC). 
Also, similar to Figure \ref{fig:selectionEffect}(b), 
it favors regions with higher ecliptic latitude, 
where {\it Gaia} observes stars more frequently (see footnote \ref{footnote:Gaia_scan})
and thus the typical parallax precision is consequently better 
(see Section \ref{sec:skyDistribution}).

Figure \ref{fig:selectionEffect}(d) shows the distribution of our initial sample of 1743 stars (see Section \ref{sec:sample}). 
These stars are selected not only because their $\vlos$ is measured by {\it Gaia} and they have high-precision parallax, but also because they have large tangential velocity, $v_{\rm tan}>300 \kms$. 
In panel (d), we see an asymmetric distribution of stars similar to that in panels (b) and (c). 
The selection in $v_{\rm tan}$ is not expected to create asymmetric distribution of stars at $b>0^\circ$ and $b<0^\circ$ at a given value of $\ell$, so we regard this asymmetric distribution as a result of our cut in $\varpi/\delta\varpi$. 
Because we see a similar pattern on the sky in Figure \ref{fig:sky}, we infer that the inhomogeneous distribution of our extreme velocity stars ($N=30$) seems to arise from our cut in $\varpi/\delta\varpi$.

\section{ADQL queries}

\subsection{Query for our initial sample}

The 1743 stars mentioned in Section \ref{sec:sample} 
can be obtained from {\it Gaia} archive 
by running the following ADQL script. 
\begin{lstlisting}
SELECT * FROM gaiadr2.gaia_source
WHERE
parallax_over_error > 10.
AND
radial_velocity is not null
AND
power(  11.1  *(-sin(radians(l))) 
      + 232.24*cos(radians(l))
      + 4.74047/parallax/cos(radians(b))
      * ((0.455984*cos(radians(dec)) 
      - 0.889988*sin(radians(dec))*cos(radians(ra-192.85948)))*pmra
      + (0.889988*sin(radians(ra-192.85948)))*pmdec)
      , 2)
+
power(  11.1  *(-cos(radians(l))*sin(radians(b))) 
      + 232.24*(-sin(radians(l))*sin(radians(b))) 
      + 7.25  *(cos(radians(b)))
      + 4.74047/parallax/cos(radians(b))
      * ((0.455984*cos(radians(dec))
      - 0.889988*sin(radians(dec))*cos(radians(ra-192.85948)))*pmdec
      + (0.889988*sin(radians(ra-192.85948)))*(-pmra))
      , 2)
> 90000.
\end{lstlisting}

\subsection{Queries for checking selection effect}

In generating Fig \ref{fig:selectionEffect}(c), 
we use the following queries. 
First, the number density of {\it Gaia} sample stars with 
measured $\vlos$ is obtained by:
\begin{lstlisting}
SELECT gaia_healpix_index(6, source_id) AS healpix_6,
count(*) / 0.8392936452111668 AS sources_per_sq_deg
FROM gaiadr2.gaia_source
WHERE radial_velocity is not null
GROUP BY healpix_6
\end{lstlisting}
Second, the number density of {\it Gaia} sample stars with 
measured $\vlos$ and good parallax ($\varpi/\delta\varpi > 10$) 
is obtained by:
\begin{lstlisting}
SELECT gaia_healpix_index(6, source_id) AS healpix_6,
count(*) / 0.8392936452111668 AS sources_per_sq_deg
FROM gaiadr2.gaia_source
WHERE radial_velocity is not null 
AND parallax_over_error>10
GROUP BY healpix_6
\end{lstlisting}

\end{document}